\documentstyle[11pt,aasms5_welty,epsfig]{article}
\newcommand{\bx}{{\bf x}}
\newcommand{\bov}{{\bf v}}
\lefthead{Zank and Frisch}
\righthead{Solar Galactic Environment}

\received{18 November 1998}
\accepted{23 December 1998}

\begin{document}
\title{Consequences of a Change in the Galactic Environment of the Sun}

\author{G.P.\ Zank}
\affil{Bartol Research Institute, University of Delaware, Newark, DE 19716}

\and

\author{P.C.\ Frisch\altaffilmark{1}}
\affil{Department of Astronomy and Astrophysics, University of Chicago, IL 60637}
\altaffiltext{1}{Currently at the University of California, Berkeley
Astronomy Department, Berkeley, California  94720-3411}

\date{}


\begin{abstract}
The interaction of the heliosphere with interstellar clouds has attracted
interest since the late 1920's, both with a view to explaining apparent
quasi-periodic climate ``catastrophes'' as well as periodic mass extinctions.
Until recently, however, models describing the solar wind - local interstellar
medium (LISM) interaction self-consistently had not been developed. Here, we 
describe
the results of a two-dimensional (2D) simulation of the interaction between 
the heliosphere and an interstellar cloud with the same properties as currently,
except that the H$^{\rm o}$ density is increased from the present 
value of $n$(H$^{\rm o}$)$\sim$0.2 cm$^{-3}$ to
10 cm$^{-3}$. The mutual interaction of
interstellar neutral hydrogen and plasma is included. The heliospheric cavity
is reduced considerably in size (approximately 10 -- 14 au to the termination
shock in the upstream direction) and is highly dynamical.  The interplanetary
environment at the orbit of the Earth changes markedly, with the
density of interstellar H$^{\rm o}$ increasing to $\sim$2 cm$^{-3}$.  The termination shock
itself experiences periods where it disappears, reforms and disappears again. 
Considerable mixing
of the shocked solar wind and LISM occurs due to Rayleigh-Taylor-like
instabilities at the nose, driven by ion-neutral friction.  Implications
for two anomalously high concentrations of ${}^{10}$Be found in Antarctic ice 
cores 33 kya and 60 kya, and the absence of prior similar events,
are discussed in terms of density enhancements in the surrounding
interstellar cloud.
The calculation presented here supports past speculation that the galactic environment of the Sun
moderates the interplanetary environment at the orbit of the Earth,
and possibly also the terrestrial climate.
\end{abstract}


\section{Introduction}

The solar system today is embedded in a warm low density interstellar
cloud (T$\sim$7000 K, n(H$^{o}$+H$^{+}$)$\sim$0.3 cm$^{-3}$), which flows 
through the solar system with a relative Sun-cloud velocity of $\sim$26
km s$^{-1}$.  Neutral interstellar gas penetrates the charged solar wind
of the heliosphere
\footnote{The heliopause bounds the heliosphere, which is the region of space
occupied by the solar wind with a radius of roughly 100 pc minimum.} --
98\% of the diffuse material in the heliosphere is interstellar gas,
and the densities of neutral interstellar gas and the solar wind are
equal at approximately the orbit of Jupiter.  The galactic environment of the 
Sun is regulated by the properties of the interstellar
cloud surrounding the solar system.
\footnote{The interstellar cloud surrounding the heliosphere is sometimes 
referred to as the ``local interstellar cloud'', or LIC.}
However, when the surrounding cloud is of low density, the solar wind prevents 
most interstellar gas and dust from reaching 1 au, the location of the Earth. 
The discovery of small scale structure with column densities 
$\geq 3$ 10$^{18}$ cm$^{-2}$
in cold interstellar matter 
(\cite{dieter,diamond,frail,meybla,watmey,heiles97}),
and the structured nature of the interstellar cloud surrounding the solar
system, allow the possibility that the spatial density of the interstellar
cloud surrounding the solar system may change within the next 10$^{4}$--10$^{6}$
years (Frisch 1995,1997a,1997b,1998; hereafter referred to as FR).
Over the past century, many conjectures have appeared
in the scientific literature linking encounters with dense interstellar clouds to possible
climate changes on Earth (e.g.\ Shapely 1921; McCrea 1975;
Begelman and Rees 1976; Fahr 1968; Reid et al. 1976; McKay and Thomas 1978; 
Scoville and
Sanders 1986; Thaddeus 1986; \cite{fr93}; \cite{Bzowski}, \cite{fr98}).  
For these suggestions
to have substance, however, it must first be shown that the
interplanetary environment of the Earth varies with changing properties
of the surrounding interstellar cloud.
It has been shown that in the past, the galactic environment of the Sun has
changed as a function of time, and that the
cloud complex sweeping
past the Sun now has an order-of-magnitude more nearby interstellar
gas in the upwind than the downwind directions (\cite{fryk86},FR).
Therefore the sensitivity of the heliosphere to variations in the boundary
conditions imposed by the LISM justify closer examination.
It is the purpose of this
paper to show that even a moderate alteration in the density of the
cloud surrounding the solar system can yield substantial variations to the
interplanetary environment in the inner heliosphere.

Early studies investigating a heliosphere embedded in a dense interstellar
cloud considered the relative ram pressures of the solar wind and surrounding
interstellar cloud to estimate the heliopause
location (e.g. \cite{holzer,holzer89}).
Contemporary models consider the interaction of 
the solar wind and interstellar medium (ISM) 
self-consistently, by including the effects of resonant charge exchange
between the ionized and neutral gases.
In the supersonic solar wind itself, charge-exchange can lead to a significant
deceleration of the wind due to the freshly ionized interstellar neutrals
extracting momentum from the solar wind. The concomitant reduction in solar
wind ram pressure can lead to a significant reduction in the size of
the heliospheric cavity.
In the boundary region separating the solar wind from
the ISM (the ``heliosheath''), neutral hydrogen charge exchange with decelerated interstellar
plasma acts to partially divert, heat and filter the H$^{\rm o}$ before it enters
the heliosphere. This filtration of H$^{\rm o}$ in the heliosheath can
reduce the number density of inflowing H$^{\rm o}$ by almost half. The rather
complicated nonlinear coupling of plasma and H$^{\rm o}$ in the vicinity of a
stellar wind is now captured in modern models (\cite{baranov}, Zank et al.,
1996a, or see e.g., Zank 1998a for a
review). The weak coupling of neutral hydrogen gas and plasma via resonant 
charge exchange
affects both distributions in
important ways. This implies that the self-consistent coupling of plasma and
neutral hydrogen is necessary
for modelling the interaction of the solar wind with the ISM. 
We employ a self-consistent two-dimensional (2D) numerical
simulation to evaluate heliospheric structure and properties 
when the heliosphere is embedded in a neutral interstellar cloud whose number
density is some thirty times greater than at present.

\begin{table*}
\caption{Parameters for Current and Modelled LISM\tablenotemark{1}}
\begin{center}
\begin{tabular}{lcccc}
& & \\ 
\hline   \\
& ``LIC''& Solar Wind & Model& Model \\
& & & Plasma  & Neutral H  \\
& ISM & (1 au) & ISM & ISM \\ 
\tableline
$n({\rm H}^{\rm o})$ (cm${}^{-3}$) &0.2&     &     & 10  \\
$n({\rm p}^{\rm +})$ (cm${}^{-3}$) &0.1--0.25& 5.0 & 0.1 & \\
$n({\rm e}^{\rm -})$ (cm${}^{-3}$) &0.1--0.25& 5.0 & 0.1 &  \\
$u$ (km s$^{-1}$) &26& 400 & --26 & --26 \\
$T$(K) &7000& $10^5$ & 8000 & 8000 \\
$M$ && 7.6 & 1.75 & 2.48 \\
\hline 
\end{tabular} \\
\tablenotetext{1}{``LIC'' refers to the current values of the LISM. 
The remaining columns list the model parameters used in the simulation.  
See Frisch et al.\ 1998 for a discussion of LIC parameters.}
\end{center}

\end{table*} 

\section{Basic Model}

A multi-fluid model is used here to model the interaction of the solar wind
with cloud of enhanced density. 
The need for modelling the neutrals as a multi-fluid
stems from the variation in the charge exchange mean-free-path for H in
different regions of the heliosphere and ISM. Large anisotropies are
introduced in the neutral gas distribution by charge exchange  with the solar
wind plasma (both sub- and supersonic regions) 
and the multi-fluid approach represents
an attempt to capture this characteristic in a tractable and computationally
efficient manner.  We consider the interaction of the heliosphere with
an interstellar cloud similar to the cloud now surrounding the
solar system, but with neutral number densities increased to 10 cm$^{-3}$.
The parameters of the enhanced density cloud, and the solar
wind at 1 au, are given in Table 1.
The relatively high neutral 
density ensures that the H$^{\rm o}$
distribution is essentially collisional. Williams et al. (1997), using the
results of Dalgarno (1960), fitted the function 
\begin{equation}
\sigma_{HH} = 3.2 \times 10^{-15} E_{eV}^{-0.11} \mbox{cm}^2 \qquad 0.1 <
E_{eV} < 100 \label{eq:3.6}
\end{equation}
to describe the cross-section for H$^{\rm o}$-H$^{\rm o}$ collisions ($E_{eV}$ is the neutral
atom energy in electronvolts). The collisional mean-free-path for H$^{\rm o}$
with the given input parameters ranges from less than 2 au in the ISM to less than 1
au in the heliospheric boundary regions (below) to less than 2.5 au in the
heliosphere itself. This suggests that the multi-fluid description outlined
below is suitable on scales larger than a few au for these moderate ISM densities, and this, as illustrated below, is shown to be the case.

The heliosphere-ISM environment can be described in terms of
three thermodynamically distinct regions; the supersonic solar wind (region 3),
the very hot subsonic solar wind (region 2), and the ISM itself (region 1).
Each region acts a source of secondary H$^{\rm o}$ atoms whose distribution reflects that
of the plasma distribution in the region. Accordingly, the neutral distribution
resulting from the interaction of the solar wind with the surrounding 
interstellar
cloud may be approximated by three distinct neutral components
originating from each region (Zank et al. 1996a).
Each of these three neutral components is represented 
by a distinct Maxwellian distribution function appropriate to the
characteristics of the source distribution in the multi-fluid models. This 
approximation allows the use of simpler production and loss terms for
each neutral component. The complete highly non-Maxwellian H distribution 
function is then the sum over the three components, 
\begin{equation}
f (\bx ,\bov ,t) = \sum_{i=1}^3 f_i (\bx ,\bov ,t) . \label{eq:3.1}
\end{equation} 
In principle, for each component, an integral equation must
be solved (Hall 1992; Zank et al. 1996b). 
Instead, Zank et al. use (\ref{eq:3.1})
to obtain three Boltzmann equations corresponding to each
neutral component. This is an extension of the procedure developed in 
Pauls et al. (1995). For component 1, both losses and gains in the 
interstellar medium need to be included, but only
losses are needed in the heliosheath and solar wind. Similarly for components 2
and 3. Thus, for each of the neutral hydrogen components $i$ ($i = 1,2$ or 3)
\begin{equation}
\frac{\partial f_i}{\partial t} + {\bf v} \cdot \nabla f_i = \left\{
\begin{array}{ll}  
P_1 + P_2 + P_3 - ( \beta_{ex} + \beta_{ph} ) f_i & \mbox{region i} \\ 
-( \beta_{ex} + \beta_{ph} ) f_i  & \mbox{otherwise}
\end{array}  \right. , \label{eq:3.2}  
\end{equation}
and $P_{1,2,3}$ means that the production or source term $P_{ex}$ is to be 
evaluated for the parameters of components 1, 2, or 3 respectively.
The $\beta_{ex}$ and $\beta_{ph}$ terms describe losses by either charge
exchange or photoionization respectively. 
Under the assumption that each of the neutral component distributions is 
approximated adequately by a Maxwellian, one obtains immediately from
(\ref{eq:3.2}) an isotropic hydrodynamic description for each neutral
component,
\begin{eqnarray}
\frac{\partial \rho_i}{\partial t} + \nabla \cdot \left( \rho_i {\bf u}_i 
\right) &=& Q_{\rho i} ; \label{eq:3.3} \\
\frac{\partial}{\partial t} \left( \rho_i {\bf u}_i \right) + \nabla \cdot 
\left[ \rho_i {\bf u}_i{\bf u}_i + p_i {\bf I} \right] &=& {\bf Q}_{mi} ; 
\label{eq:3.4} \\
\frac{\partial}{\partial t} \left( \frac{1}{2} \rho_i u_i^2 + 
\frac{p_i}{\gamma - 1}
\right) + \nabla \cdot \left[ \frac{1}{2} \rho_i u_i^2 {\bf u}_i +
\frac{\gamma}{\gamma - 1} {\bf u}_i p_i \right] &=& Q_{ei} . \label{eq:3.5}
\end{eqnarray}
The source terms $Q$  
are listed in Pauls et al. (1995) and Zank et al. (1996a). 
The subscript $i$ above refers to the neutral component  of interest ($i =
1,2,3$), $\rho_i$, ${\bf u}_i$, and $p_i$ denote the neutral component $i$
density, velocity, and isotropic pressure respectively, ${\bf I}$ the unit
tensor and $\gamma$ ($= 5/3$) the adiabatic index.

The plasma is described similarly by the 2D hydrodynamic equations
\begin{eqnarray}
\frac{\partial \rho}{\partial t} + \nabla \cdot \left( \rho {\bf u} \right) &=&
Q_{\rho p} ; \label{eq:3.7} \\
\frac{\partial}{\partial t} \left( \rho {\bf u} \right) + \nabla \cdot \left[
\rho {\bf u}{\bf u} + p {\bf I} \right] &=& {\bf Q}_{m p} ; \label{eq:3.8} \\
\frac{\partial}{\partial t} \left( \frac{1}{2} \rho u^2 + \frac{p}{\gamma - 1}
\right) + \nabla \cdot \left[ \frac{1}{2} \rho u^2 {\bf u} +
\frac{\gamma}{\gamma - 1} {\bf u} p \right] &=& Q_{e p} , \label{eq:3.9}
\end{eqnarray}
where $Q_{(\rho,m,e),p}$ denote the source terms for plasma density, momentum, 
and energy. 
These terms are also defined in Pauls et al. (1995) and Zank et al. (1996a). The remaining symbols enjoy their 
usual meanings. The proton and electron temperatures are assumed equal in the
multi-fluid models.

The coupled multi-fluid system of equations (\ref{eq:3.3}) -- (\ref{eq:3.9}) 
are solved numerically as described in Pauls et al. (1995) and Zank et al. (1996a).  

\section{Simulation Results}

\subsection{Global Heliosphere Configuration}

The global structure of the heliosphere embedded in a high density
environment is obtained by integrating the 
coupled time-dependent system of hydrodynamic equations
(\ref{eq:3.3}) -- (\ref{eq:3.5}) and (\ref{eq:3.7}) -- (\ref{eq:3.9}) 
numerically in
two spatial dimensions.   
The integration begins at time $t$=0 using as an initial condition the
heliosphere embedded in a low density cloud (Zank et al. 1996a). 
The LISM neutral number density is increased to 10 cm${}^{-3}$ to mimic the
``encounter'' of the heliosphere
with a denser cloud, and it is shown below that the heliosphere then fails to
settle into a steady-state or equilibrium configuration.

A time sequence of the 2D global plasma structure is
illustrated in Plate 1. 
\begin{figure}
\caption{{\bf Plate 1}: \tiny 
A time sequence of the plasma temperature distribution. The
color corresponds to the Log[Temperature]. The ordering A, B, C, D corresponds
to a temporal separation from the preceding figure of $\sim 66$ days. The four
figures show an approximately full evolutionary cycle which is repeated on a
$\sim 280$ day period. See text for further details.}
\end{figure}
The time dependence of both the plasma and neutrals 
is presented separately below.
Four successive figures are shown, each separated from
the preceding in time by $\sim 66$ days. The color depicts the plasma
temperature. The Sun is located at the origin and the interstellar wind is
assumed to flow from the right boundary (located at 1000 au; a numerical
semi-circle of radius 1000 au forms the computational domain) with the
parameters listed in Table 1. Since the plasma interstellar Mach number is
supersonic (see Zank et al.\ 1996a for a discussion about a subsonic ISM), a
bow shock, labelled BS in Plate 1A, forms, which decelerates, heats and
diverts the interstellar plasma around the heliospheric obstacle. 
The plasma is now no longer collisionally equilibrated with the interstellar
H$^{\rm o}$ and charge exchange between the interstellar plasma and
neutrals just
downstream of the bow shock leads to an effective heating/compression and
diversion of the neutral interstellar H in the heliosheath region (between
the bow shock and heliopause). As a result, a pile-up of
decelerated, diverted and heated interstellar H$^{\rm o}$ is created in the region downstream
of the bow shock, which is illustrated in Plate 2. 
\begin{figure}
\caption{{\bf Plate 2}: \tiny 
A time sequence (corresponding to that of Plate 1) of the global
distribution of the neutral interstellar H density. The color refers to the
density measured in cm${}^{-3}$. The hydrogen wall in the upstream direction is 
clearly visible, as is the effective filtration of H as it enters the
heliosphere.}
\end{figure}
Plate 2 shows the
corresponding global distribution of H$^{\rm o}$ at the same times as used in
Plate 1, except that the H$^{\rm o}$ density is plotted.  Far upstream of the
bow shock, the H$^{\rm o}$ number density is $\sim 10$ cm${}^{-3}$, increasing
rapidly just downstream of the bow shock in the pile-up region with 
densities 2--2.5 times greater than the ambient ISM density. 
Changes in the neutral number density, velocity, temperature and Mach number
are seen more easily in the one-dimensional (1D) plots along the stagnation 
axis than in Plate 2 (discussed later in the context of Figure 2). 
The number density of neutrals entering the heliosphere at $\sim
10$ au is $\sim 7$ cm$^{-3}$, decreasing to $\sim 2$ cm$^{-3}$ at 1 au.

The enhanced-density ISM case can be compared to simulations using the current 
low-density surrounding ISM (the LIC column of Table 1).  The hydrogen wall
of the high density cloud model 
possesses a more complicated structure than that 
of the low density case -- it possesses a weak double-peaked 
structure along the stagnation axis in the upstream direction,
reverts to a single-peaked structure away from but in the neighbourhood of the
stagnation axis, after which it again becomes double-peaked. The H$^{\rm o}$ 
flow is strongly filtered in the vicinity of the wall, this due essentially to 
the divergence of the H$^{\rm o}$ flow through charge exchange with the 
bow shock-diverted 
interstellar plasma, thus leading to an H$^{\rm o}$ number density entering
the heliosphere that is much lower than that of the ISM.

As is well-known (e.g. Holzer 1972), the effect of interstellar H streaming
through the supersonic solar wind, and experiencing charge exchange with solar
wind protons, is to reduce the momentum of the solar wind significantly. The
relatively large H$^{\rm o}$ number density entering the heliosphere 
reduces the solar wind ram pressure significantly, 25\%--75\%, 
from its nominal non-mediated value thereby reducing the global extent of the 
heliosphere. For the parameters of Table 1, the
heliosphere shrinks dramatically to $\sim 10$ -- 14 au in the upstream
direction.  The density of interstellar neutral hydrogen at 1 au
becomes $\sim$ 2 cm$^{-3}$ (Figure 2), unlike the current state for which 
no neutral interstellar hydrogen reaches Earth orbit 
(e.g. \cite{baranov}; Zank et al., 1996a).  
In addition, the configuration becomes unstable (see below).

The reduction in solar wind momentum is accompanied by an increase 
in the total temperature of the solar wind (solar wind ions plus pickup ions),
in comparison to an adiabatically expanding solar wind, since pickup
ions acquire a large velocity perpendicular to both the solar wind velocity
and magnetic field vectors and therefore have energies that are
typically $\sim 1$ keV in the unmediated solar wind (where $u \sim 400$ km s$^{-1}$).
Accordingly, beyond the ionization cavity (a cavity 
created by the photoionization of
interstellar neutrals by solar wind UV radiation), the solar wind temperature
increases (in the absence of pickup ions, the total solar wind
temperature decays adiabatically, the polytropic index depending on the nature
of the specific heat sources identified to heat the wind).
Radial cuts along the stagnation axis for the plasma
density, velocity, temperature, and Mach number are illustrated in Figure 1.
\begin{figure}
{\epsfig{figure=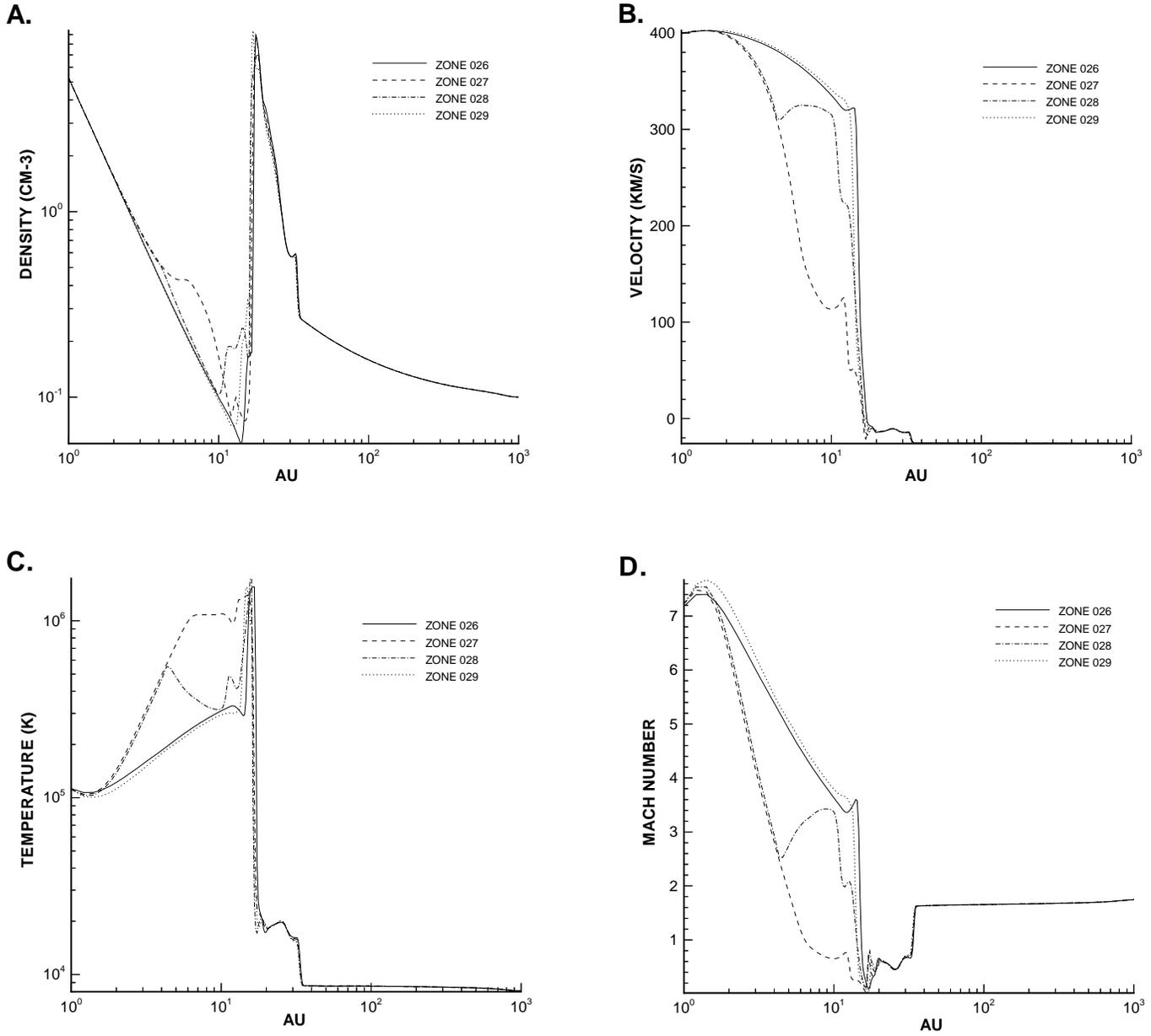,width=15cm,height=20cm}}
\caption{{\bf Figure 1}: \tiny 
The plasma density (A), velocity (B), temperature (C) and Mach
number (D) along the stagnation axis. Four curves are presented on each graph,
each corresponding to a cut through one of the figures of Plate 1. In each
case, the solid line corresponds to Plate 1A, the dashed line to Plate 1B, the
dash-dotted line to Plate 1C and the dotted line to Plate 1D.}
\end{figure}
The increase in the solar wind temperature is evident in Figure 1c.  By 
contrast, in the
absence of pickup ions or other heating processes, the solar wind temperature
would cool adiabatically according to $r^{-4/3}$ for an adiabatic index of
5/3.
\begin{figure}
{\epsfig{figure=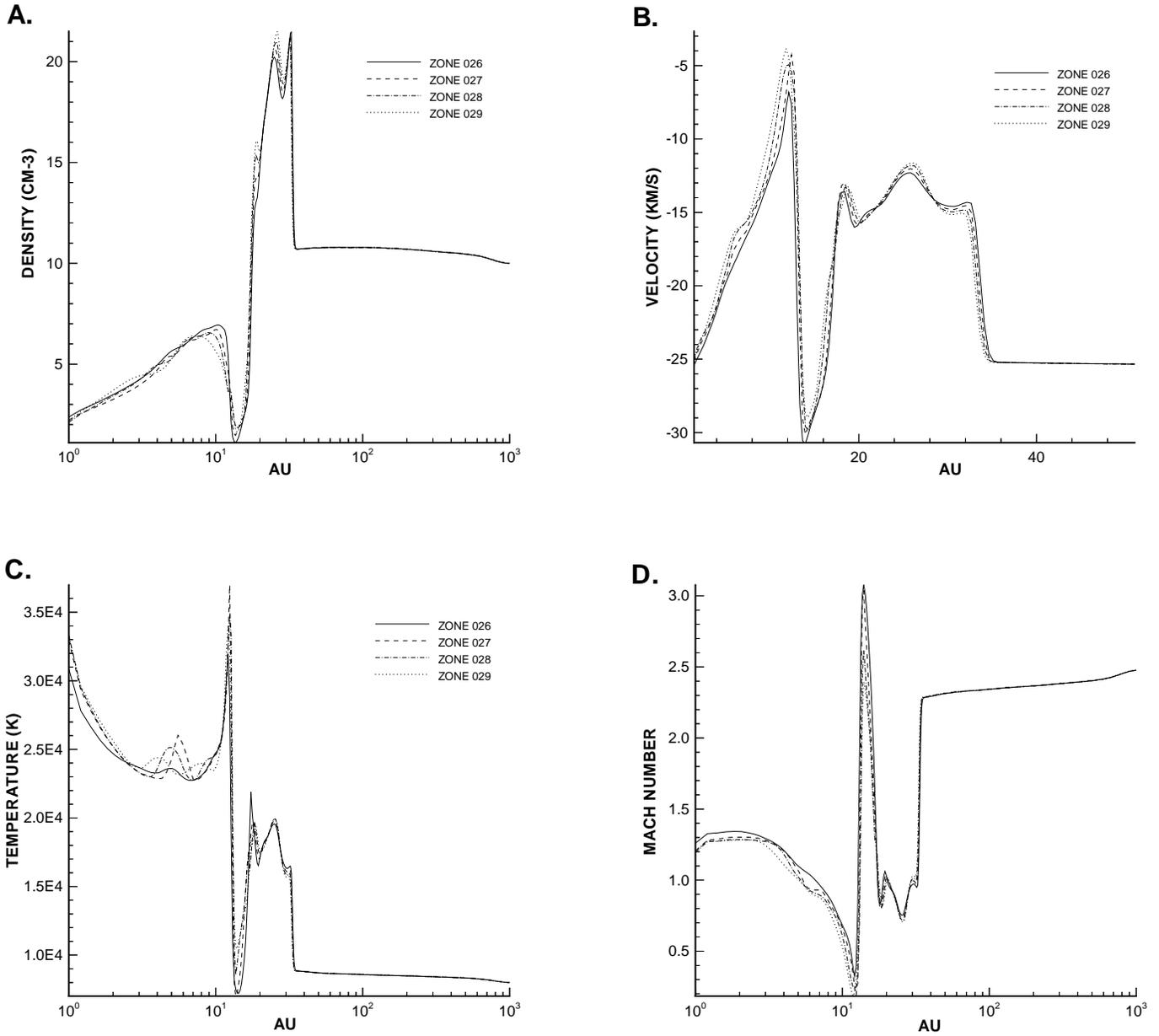,width=15cm,height=20cm}}
\caption{{\bf Figure 2}: \tiny 
The interstellar neutral density (A), velocity (B), temperature
(C) and Mach number (D) along the stagnation axis. Four curves are presented 
on each graph,
each corresponding to a cut through one of the figures of Plate 2. In each
case, the solid line corresponds to Plate 2A, the dashed to Plate 2B, the
dash-dotted to Plate 2C and the dotted to Plate 2D.}
\end{figure}

\subsection{Time Dependence}

The very high number density of interstellar neutrals entering the
heliosphere, even after filtration ($\sim 7$ cm${}^{-3}$, Plate 2, Figure 2a)
leads to a very strong time-dependent mediation of the solar wind.  
In the sequence of velocity
profiles plotted in Figure 1b (the sequence corresponds to the four plot
sequence of Plates 1 and 2), the solar wind at the time corresponding to the
solid line is decelerated from 400 km s$^{-1}$ to $\sim 320$ km s$^{-1}$ at the termination
shock. Some 66 days later (the dashed line), the deceleration is considerably
larger, now down to $\sim 110$ km s$^{-1}$. 
Neutrals that were ionized and then picked up by the solar wind during the
least mediated solar wind period will therefore have considerably more energy
than those picked up by a strongly mediated wind (a factor $\sim 10$
difference). Thus, the strongly mediated solar wind will possess a very large
temperature gradient. The temperature gradient can become so large that the
pressure gradient, which normally decreases with increasing radial distance in 
weakly mediated solar wind, is now inward (Figure 3), and it is this inwardly
directed pressure gradient that is responsible for decelerating the outwardly
flowing solar wind so strongly. As a result of the much stronger deceleration
of the solar wind, the radial solar wind density profile departs 
from an $\sim r^{-2}$ dependence (Figure 1a) on occasion.
\begin{figure}
{\epsfig{figure=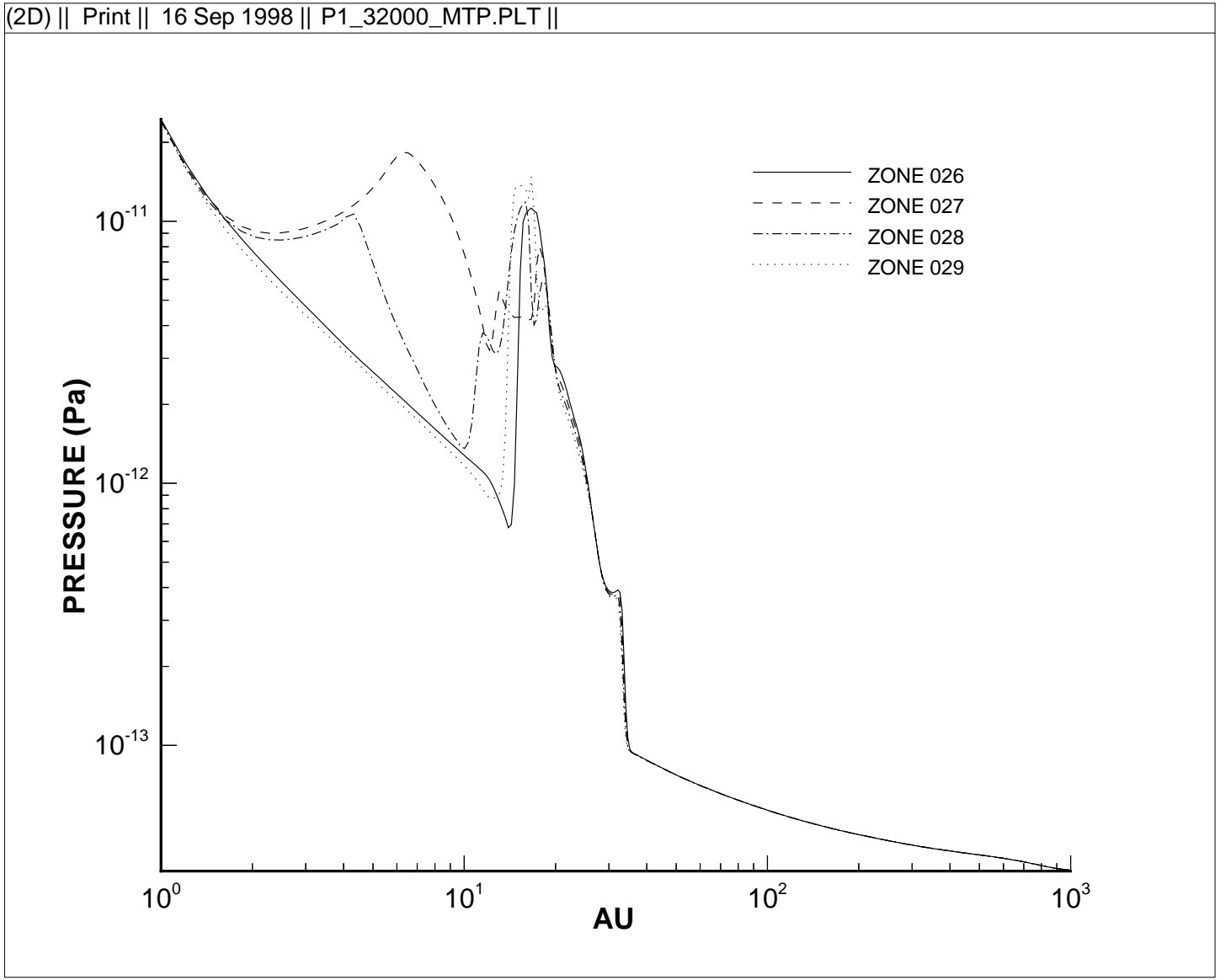,width=15cm,height=20cm}}
\caption{{\bf Figure 3}: \small
The plasma pressure along the stagnation axis. As in Figure 1,
four curves are plotted, each of which corresponds to a figure from Plate 1 and
the same line convention is followed as in Figure 1.
}
\end{figure}

Once the solar wind is mediated strongly by pickup ions (the dashed curve,
Figure 1), the newly born ions are much less energetic and no longer contribute
as strongly to the temperature and pressure build-up in the outer heliosphere.
Consequently, the outer heliosphere begins to cool, the inwardly directed
pressure gradient is reduced, and the solar wind begins to revert to a less
strongly mediated state. In the simulation, the periodic weak and strong
mediation of the solar wind is very evident when the results are plotted
in a movie format, and temperature fronts are seen to propagate continually out
of the solar wind. This can be seen in the temperature sequence of Plate 1. 
The numerical simulations have been run for very long times, but no abatement
of the temporal behaviour is observed. Essentially, the number density of
neutrals entering the heliosphere is far too large to allow the solar wind to
ever settle into a steady or quasi-steady state. 
The large neutral number density forces the wind to deviate 
strongly from its zeroth-order unmediated
state, which in turn drives strong variations
in the heating effects introduced by pickup ions.

Associated with the periodic increase of plasma temperature in the outer
heliosphere is a decrease of the solar wind Mach number. The Mach number, which
decreases in a pickup ion mediated solar wind because of both the solar wind
deceleration and the increased sound speed, normally remains larger than 1
(supersonic). However, the strongly mediated solar wind Mach number decreases
to 1 and less (Figure 1d). In this case, a termination shock is no longer
necessary to bound the solar wind and a shock-free transition from supersonic
to subsonic flow occurs (Plate 1b). As the outer heliospheric temperature begins
to decrease again, the Mach number increases and a termination shock is again
present. The sequence showing the presence of a termination shock, then a
shock-free transition, followed by the reformation of the termination shock is
exhibited in Plate 1A, B, and C respectively.

Finally, it should be noted that the heliopause region is unstable to a
Rayleigh-Taylor-like instability. This is discussed in more detail in Zank
(1998b)
but the effect is understood easily. The coupling of neutral interstellar H to
plasma introduces a frictional term proportional to $\langle \sigma v \rangle N
\rho ({\bf u} - {\bf V}_n )$ in the momentum balance, 
where $\langle \sigma v \rangle$ denotes the
charge exchange cross-section, $N$ the neutral number density, ${\bf V}_n$ the
neutral velocity, and $\rho$ and ${\bf u}$ the plasma variables
density and velocity. This
expression is somewhat simplified compared to the full expressions used in the
simulation. The inwardly directed frictional term plays a similar role as the 
gravitational term in the classical Rayleigh-Taylor analysis (e.g.
Chandrasekhar 1961) and the charge-exchange ion-neutral friction term can
therefore destabilize the heliopause. This leads to a somewhat unstable flow
pattern in the upstream region, and some mixing of the solar wind and
interstellar plasma.

In Plate 2, the global neutral distribution is also time-dependent but much
more weakly than the plasma. Plate 2 suggests that the overall structure of
the heliospheric hydrogen wall does not change very significantly with time.

\section{Discussion and Conclusions}

It is shown in this paper, using a coupled time-dependent system of hydrodynamic equations
to describe the interaction between the heliosphere and a partially
ionized interstellar cloud, that the configuration of the heliosphere and the
interplanetary environment of the Earth are sensitive to the
density of neutrals in the interstellar cloud surrounding the Sun.
Raising the neutral number density in the surrounding cloud
to $\sim 10$ cm$^{-3}$, and leaving the other parameters at current values,
has a dramatic effect on the global spatial and temporal structure of the
heliosphere. The heliosphere in the upstream direction is reduced 
in extent to $\sim$10--14 au, compared to $\sim 80$ - $\sim$120 au currently. 
Interstellar H$^{\rm o}$ is filtered considerably as it enters the
heliosphere, but some survives to penetrate successfully to 1 au where
interstellar H$^{\rm o}$ densities are $\sim$2 cm$^{-3}$, 
unlike the current
conditions for which interstellar hydrogen is ionized entirely a few
au from the Sun.

Although we have changed only the density of the surrounding cloud in these
simulations, denser interstellar clouds are expected to have somewhat higher
magnetic field strengths than the low value found locally. An average value of
7.3$\mu$G has been found recently for neutral interstellar gas from Zeeman
splitting (\cite{lockman}). Were the interstellar magnetic field strength also
to be increased to 7$\mu$G in our simulations, we would expect the heliopause
radius to decrease possibly to a half to a third (depending on magnetic field
orientation) of our computed radius, using a very crude balance estimate. 

The heliospheric configuration becomes unstable; properties
of the solar wind in the outer heliosphere vary on timescales as short as
$\sim 66$ days, and the termination shock disappears at times.
The formation, disappearance and reformation of the termination shock
appears to be periodic with a timescale of $\sim 280$ days. Pickup ions
dominate the thermodynamical character of the outer heliospheric solar wind
completely. Solar wind temperatures, when pickup ions are included, can exceed 
$10^5$ K. An inward pressure gradient can develop, which causes the solar wind
to decelerate far more than predicted by simple mass-loading arguments and,
consequently, the density decrease with increasing heliocentric distance can
deviate from the usual $r^{-2}$ dependence.

The implications of high interstellar neutral densities 
at 1 au for the terrestrial environment are interesting 
(e.g., Begelman and Rees, 1976; \cite{Bzowski}). 
Is there evidence that the Earth has passed through
a dense interstellar cloud in the (pre-)historical past, or might do so in the
near future?  A cloud condensation with density $\sim$10 cm$^{-3}$
embedded in the interstellar cloud complex flowing past the solar
system could have a thickness of up to 0.3 pc and not violate any
known observational constraints. Such a condensation would pass over the
solar system in less than 15,000 years.
There are several sets of data that indicate that the galactic environment
of the Sun may have changed recently.  The first set are geological
records of nitrate layers $\sim$ 11000 years before present, which have been attributed to 
energetic photons released by the Vela supernova explosion (\cite{vela}; also
see \cite{frsla96}).  The second set of data is absorption
line measurements of a slightly denser and cooler interstellar cloud seen within
3 pc of the Sun towards the stars $\epsilon$ CMa and $\alpha$ CMa 
in the anti-apex direction (see the discussion of the distribution
of the nearby ISM in Frisch 1997b).  The second cloud at $\sim$--17 km s$^{-1}$
(heliocentric velocity, sometimes referred to as the ``blue-shifted'' cloud) has an electron density n(e$^{-})$$\sim$0.4 cm$^{-3}$,
and temperature T=3600 K, in comparison with the values $\sim$0.1--0.2 cm$^{-3}$
and 7000 K seen locally (\cite{grydupin,lallement}).

A third kind evidence is the cosmic ray record
implied by the spikes in the $^{10}$Be record found in Antarctic ice
core samples at ages corresponding to 33 000 and 60 000 years ago (the D1 and D2 events respectively). 
Raisbeck et al. (1987) and Sonett et al. (1987) suggest that an
increase in the cosmic ray flux on the Earth's atmosphere can lead to an
enhancement in the precipitated beryllium.
Since supernova remnant shocks are thought to be responsible for accelerating
galactic cosmic rays up to energies of $\sim 10^{14}$ eV per nucleii, Sonett et al.\
suggested that the spikes might correspond to the passage
of supernova shocks through the heliosphere. 
However, Frisch (1997a) has argued that little observational evidence in the LISM
exists for the presence of a supernova shock at the times needed to explain the
${}^{10}$Be enhancements. We suggest below
an alternative interpretation of the ${}^{10}$Be enhancements. 

There are two possible ways that the
cosmic ray flux at 1 au can be increased by the interaction of the solar wind
with a dense interstellar cloud. The first is that the increased pickup ion
population increases the anomalous cosmic ray population.  The second is that
the reduction in the size of the heliospheric cavity indicates that
galactic cosmic rays are no longer modulated significantly
by an extended solar wind and
so, at 1 au, the Earth samples the full spectrum. A more quantitative
connection between the anomalous cosmic ray flux and a high density neutral
interstellar cloud can be made using the model discussed above.
The pickup ion number density at the termination shock, located at $\sim
10$ -- 14 au, may be estimated as $n_{PI} (10\mbox{au} ) \simeq 3 \times
10^{-2}$ cm$^{-3}$ (after solving $1/r^2 \; d/dr (r^2 u n_{PI} ) = \langle
\sigma v \rangle N_H n_{SW}$, under the assumptions that the solar wind number
density $n_{SW} \simeq n_0 (r_0 /r )^2$ on average, $\langle \sigma v \rangle
\simeq 10^{-15}$ cm$^2$, and using Figure 2a). The injection of pickup ions
into the anomalous cosmic ray component at a quasi-perpendicular termination
shock has been considered by Zank et al. (1996c) and Lee et al. (1996). On the
basis of such a shock ``surfing'', or multiply reflected ion acceleration mechanism,
Zank et al. estimated an injection efficiency of $\sim 20${\%} for
hydrogen, with correspondingly smaller values for the heavier pickup ions. 
Thus, the number density of anomalous cosmic rays at 10 au for a heliosphere
embedded in a H$^{\rm o}$ interstellar cloud of density 10 cm$^{-3}$ is $\sim 6
\times 10^{-3}$ cm$^{-3}$. To determine the cosmic ray density at 1 au, we
can use a simple convection-diffusion solution to the cosmic ray transport
equation, $n_{CR} (r \mbox{au} ) = n_{CR} (10 \mbox{au} ) \exp \left[ -u (10 -
r) / \kappa_{rr} \right] $ where $\kappa_{rr}$ is the radial diffusion
coefficient for cosmic rays in the heliosphere. Using an estimate 
for the radial mean free path $\lambda_{rr} \sim 0.01$ au for
100 MV 
rigidity ions (Zank et al. 1998) yields $n_{CR} (1 \mbox{au} ) 
\sim 10^{-4}$ cm${}^{-3}$. 
This is a very high level indeed. With such a large flux of anomalous cosmic
rays incident on the atmosphere of the Earth, one might therefore expect a 
noticeable
enhancement in the ${}^{10}$Be flux measured on Earth. This effect will
be further enhanced by the reduced modulation of galactic cosmic rays in the smaller
heliosphere.  

This alternative explanation for the $^{10}$Be spikes offers a natural
explanation for the absence of ${}^{10}$Be enhancements prior to the
D2 event in terms of the recent history of the galactic environment of the Sun
(\cite{fr97a,fr98}).
During those periods when the heliosphere is embedded in
an interstellar cloud with appreciable densities of H$^{\rm o}$, the anomalous
and galactic cosmic ray fluxes incident on the Earth's atmosphere will
increase and enhance the level of precipitated ${}^{10}$Be. 
During the time prior to the D2 event (60 000 years ago), the Sun was
in the third-quadrant void where the heliosphere would have been at least as 
large as today (\cite{fr98}), entering the Local Fluff cloud complex
only within the last $\sim 10^5$ years.

The effect of a large population of
H$^{\rm o}$ atoms filling interplanetary space at 1 au may well 
affect the interaction of the terrestrial magnetosphere with the
solar wind, since interstellar He$^{\rm o}$ densities will increase to
$\sim$1 cm$^{-3}$ at 1 au for this increased density case and be comparable
to the density of interstellar H$^{\rm o}$ atoms.   In addition,
enhanced cosmic ray fluxes at 1 au may alter the global electric circuit, since
the cosmic ray flux is the dominant source of conductivity in the
lower atmosphere (\cite{roble});  the global electrical circuit has
been postulated to play a role in the terrestial climate 
(Tinsley 1994,1997).


\noindent
\acknowledgments
GPZ is supported in part by an NSF Young Investigator
Award ATM-9357861, an NSF award ATM-9713223, a NASA award
NAG5-6469, a NASA Delaware Space Grant NGT5-40024, and JPL contract 959167. PCF
acknowledges the support of NASA grants NAG 5-6405 and NAG 5-7007.

\clearpage



\clearpage

\section*{Figure Captions}

{Plate 1:  A time sequence of the plasma temperature distribution. The
color corresponds to the Log[Temperature]. The ordering A, B, C, D corresponds
to a temporal separation from the preceding figure of $\sim 66$ days. The four
figures show an approximately full evolutionary cycle which is repeated on a
$\sim 280$ day period. See text for further details. \label{plate1}}

{Plate 2:  A time sequence (corresponding to that of Plate 1) of the global
distribution of the neutral interstellar H density. The color refers to the
density measured in cm${}^{-3}$. The hydrogen wall in the upstream direction is 
clearly visible, as is the effective filtration of H as it enters the
heliosphere. \label{plate2}}

{Figure 1: The plasma density (A), velocity (B), temperature (C) and Mach
number (D) along the stagnation axis. Four curves are presented on each graph,
each corresponding to a cut through one of the figures of Plate 1. In each
case, the solid line corresponds to Plate 1A, the dashed line to Plate 1B, the
dash-dotted line to Plate 1C and the dotted line to Plate 1D. \label{fig1}}

{Figure 2: The interstellar neutral density (A), velocity (B), temperature
(C) and Mach number (D) along the stagnation axis. Four curves are presented 
on each graph,
each corresponding to a cut through one of the figures of Plate 2. In each
case, the solid line corresponds to Plate 2A, the dashed to Plate 2B, the
dash-dotted to Plate 2C and the dotted to Plate 2D. \label{fig2}}

{Figure 3: The plasma pressure along the stagnation axis. As in Figure 1,
four curves are plotted, each of which corresponds to a figure from Plate 1 and
the same line convention is followed as in Figure 1.  \label{fig3}}

\end{document}